\documentclass[12pt,english]{article}
\usepackage{epsf}
\usepackage{geometry}
\usepackage{setspace}
\usepackage{amssymb}

\makeatletter

\usepackage{setspace}

\makeatother
\begin{document}

\title{\textbf{\large Integral equations in MHD: theory and application}}

\renewcommand{\thefootnote}{\fnsymbol{footnote}}
\author{F. STEFANI\footnote{Corresponding author. Email: F.Stefani@hzdr.de} $\dagger$, 
M. XU $\ddagger$, 
G. GERBETH$\dagger$, and
T. WONDRAK$\dagger$ }

\maketitle

\begin{center}$\dagger${\small Helmholtz-Zentrum Dresden-Rossendorf, P.O. Box 510119,
D-01314 Dresden,
Germany}\\
$\ddagger${\small Shandong University, P. O. Box 88, Jing Shi Road 73 , Jinan City, Shandong Province, P. R. China}
\end{center}

{\footnotesize
The induction equation of kinematic magnetohydrodynamics is mathematically
equivalent to a  system of integral equations for the magnetic field
in the bulk of the fluid and for the
electric potential at its boundary. We summarize the
recent developments concerning the numerical implementation of this
scheme and its applications to various forward and inverse problems in
dynamo theory and applied MHD.
}

\noindent \textit{\footnotesize Keywords:} {\footnotesize Dynamo;
integral equations; inverse problems
}{\footnotesize \par}

\section{Introduction}

This paper is intended as a summary and update of our recent work 
on the theoretical formulation, the numerical implementation, 
and the practical application of the integral equation approach (IEA)
to kinematic magnetohydrodynamics (MHD). 
Actually, our activity in this field had started with the 
paper (Stefani {\it et al.}, 2000), written together with 
Karl-Heinz R\"adler, whom we would like 
to thank not only for his insistence on a rigorous 
mathematical formulation in this particular case,
but also for many fruitful discussions over the course of time.

The theory of hydromagnetic dynamos 
is mainly concerned with the self-excitation of cosmic magnetic 
fields, with  particular focus on the fields of planets, stars, and 
galaxies (Krause and R\"adler 1980). 
As long as the self-excited magnetic field is 
weak and its influence on the velocity field is 
negligible we speak about the {\it kinematic dynamo regime}. When the 
magnetic field has grown to higher amplitudes the field-generating
velocity field is modified, and the dynamo enters its 
{\it saturation regime}, which
will not be considered in the present paper, though. 

The familiar way to deal with kinematic dynamo action relies on
the induction equation for the magnetic field $\textbf{B}$,
\begin{eqnarray}{\label{eq1}}
\frac{\partial \textbf{B}}{\partial t}=\nabla
\times(\textbf{u}\times\textbf{B})+\frac{1}{\mu_0\sigma}
\Delta \textbf{B},\;\;\nabla\cdot \textbf{B}=0,
\end{eqnarray}
where $\textbf{u}$ is considered as a pre-given velocity field, 
$\mu_0$ is the magnetic permeability constant, and 
$\sigma$ is electrical 
conductivity of the fluid. The time dependence of the magnetic field $\textbf{B}$ in
Eq. (\ref{eq1}),
in terms of exponential
growth or decay, is governed 
by the ratio of field changes due to 
velocity gradients (first term on the r.h.s.) to the field dissipation 
(second term on the r.h.s.). This ratio
is determined by the magnetic Reynolds number 
$R_m=\mu\sigma LU$, where 
$L$ and $U$ are typical length and velocity 
scales of the flow, respectively. 
When the magnetic Reynolds number reaches a 
critical value, denoted by $R_m^c$, the field can grow exponentially 
in time.

Equation (\ref{eq1}) follows directly from pre-Maxwell's 
equations and Ohm's law in 
moving conductors. In order to make this equation solvable, 
boundary conditions of the 
magnetic field must be 
prescribed. 
In the case of vanishing excitations of the magnetic field from 
outside the considered 
finite region, the boundary condition of the magnetic field is 
given by $\textbf{B}=O(r^{-3})\; \mbox{as}\; r\rightarrow\infty$. 
For dynamos in spherical domains, as they are typical for 
planets and stars, the problem of implementing the 
non-local boundary 
conditions for the magnetic field is easily solved by using decoupled 
boundary conditions for each degree of the spherical harmonics.
For other than spherically
shaped dynamos, in particular for the recent laboratory dynamos
working often in cylindrical domains, but also for 
dynamos in galaxies and planetesimals,  
the handling of the  non-local boundary 
conditions is a notorious problem (Stefani {\it et al.} 2009). 

A simple, but rather rude way to circumvent this 
problem is to replace the complicated non-local boundary 
conditions by simplified
local ones (so-called ''vertical field conditions''). This method is
sometimes used in the simulation of galactic dynamos
(a recent example is Hubbard and Brandenburg 2010). 

For the simulation of the cylindrical 
Karlsruhe dynamo experiment, R\"adler {\it et al.} (1998, 2002)
had used the alternative trick of embedding 
the actual electrically conducting dynamo domain into a sphere, and the region 
between this
sphere and the surface of the dynamo was virtually filled by a medium of 
lower electrical conductivity. By reducing the value of this conductivity,
it was possible to check the convergence of the numerical solution.

Of course, both methods are connected with losses of accuracy, 
the latter method being  certainly more accurate. In order to 
fully implement the non-local 
boundary condition, Maxwell's equations 
must be fulfilled in the exterior, too. 
This can be implemented in different ways.
For the finite difference simulation of the Riga dynamo, 
a Laplace equation was solved, for each time-step, by a 
pseudo-relaxation method, in the
exteriour of the dynamo domain and the magnetic field solutions in 
the interiour and in
the exterior were matched using interface conditions
(Stefani {\it et al.} 1999, Kenjere\v{s} {\it et al.} 2006).
A similar method, although based on the finite element method,  
was developed by Guermond et al. (2003, 2007)
and was intensively used later for practical dynamo applications 
(Guermond {\it et al,} 2009, Giesecke {\it et al.} 2010a, b,
Nore {\it et al.} 2011). 

A  quite elegant technique to circumvent the solution in 
the exteriour  was presented  by Iskakov {\it et al.} (2004, 2005)
who used a  combination of a finite volume and a boundary element method.
This latter method was recently used by Giesecke {\it et al.} 
(2008, 2010a, b)
for the simulation of the French VKS-dynamo (although the main
focus of that work laid on the particular effect of the 
high magnetic permeability impellers on the mode selection in this
type of dynamo).

An  alternative to the solution of the induction equation is the 
integral equation approach (IEA)
for kinematic dynamos which basically relies on the self-consistent 
treatment of Biot-Savart's law. 
For steady 
dynamo action in infinite domains of homogeneous conductivity, the
IEA  becomes very simple and had already been 
employed by a few  authors (Gailitis 1970, Gailitis and Freibergs 1974,
Dobler and R\"adler 1998). 
For dynamo action in finite domains, however, the simple Biot-Savart equation 
has to be supplemented by a boundary 
integral equation for the electric potential (Roberts 1967, 
Stefani {\it et al.} 2000, Xu {\it et al.} 2004a). If the 
magnetic field becomes time-dependent, yet another equation for the 
vector potential can be added to get a closed system of integral 
equations (Xu {\it et al.} 2004b). 

In the following section, we will present this formulation
and its specifications in detail. Then we will consider its 
application to various dynamo 
problems, including those with different degrees of symmetry which enable 
dimensional reductions.
At last, we will go over from forward problems to inverse problems
for which the IEA represents an ideal
mathematical starting point.

\section{Mathematical formulation}

Assume a fluid of electrical conductivity $\sigma$, flowing with 
velocity $\bf u$, to 
be confined in a finite region $V$ 
with boundary $S$, the 
exterior of this region consisting of insulating material (or vacuum).
By further assuming the velocity $\bf u$ field to be stationary,
we may write the electric field, the magnetic field and the magnetic
vector potential in the following form:
\begin{eqnarray}
{\bf E}({\bf r},t)={\bf E}({\bf r})\exp{\lambda t}, 
{\bf B}({\bf r},t)={\bf B}({\bf r})\exp{\lambda t},
{\bf A}({\bf r},t)={\bf A}({\bf r})\exp{\lambda t}, {\label{eq2}}
\end{eqnarray}
where the real part of $\lambda$ is the 
growth rate, and its imaginary part is the frequency of the fields.
In the general case, we allow
external currents to be present that produce a magnetic field
${\bf B}_0$. The remaining magnetic field, induced  by the
flow, is denoted by $\bf b$. 
Then, the current density is governed by Ohm's law 
\begin{eqnarray}{\label{eq3}}
{\bf j}&=&\sigma ({\bf E} + {\bf u} \times ({\bf B}_0 + {\bf b}))  \nonumber \\
&=&\sigma ( -\nabla \phi -\lambda {\bf A} + {\bf u} \times ({\bf B}_0 + {\bf b}))
\end{eqnarray}
where the electric field {\bf E}, the electric potential $\phi$ and the 
magnetic vector potential $\bf A$ are used.
Applying now Biot-Savart's law to the magnetic field, 
Green's second theorem to the solution
of the Poisson equation for the electric potential, and 
Helmholtz's theorem for the 
integral representation of the vector potential in terms of the 
magnetic field, we arrive at
the following integral equation system:
\begin{eqnarray}
{\mathbf{b}}({\mathbf{r}})&=&\frac{\mu\sigma}{4\pi}\int_V\frac{({\mathbf{u}}({\mathbf{r}}')
\times({\mathbf{B}}_0({\mathbf{r}}')+{\mathbf{b}}({\mathbf{r}}')))
\times({\mathbf{r}}-{\mathbf{r}}')}
{|{\mathbf{r}}-{\mathbf{r}}'|^3}dV' \nonumber\\
&&-\frac{\mu\sigma\lambda}{4\pi}\int_V\frac{{\mathbf{A}}({\mathbf{r}}')
\times({\mathbf{r}}-{\mathbf{r}}')}{|{\mathbf{r}}-{\mathbf{r}}'|^3}
dV'-\frac{\mu\sigma}{4\pi}\int_S\phi({\mathbf{s}}'){\mathbf{n}}({\mathbf{s}}')\times
\frac{{\mathbf{r}}-{\mathbf{s}}'}{|{\mathbf{r}}-{\mathbf{s}}'|^3}dS' \label{eq4}\\
\frac{1}{2}\phi({\mathbf{s}})&=&\frac{1}{4\pi}\int_V\frac{({\mathbf{u}}({\mathbf{r}}')
\times({\mathbf{B}}_0({\mathbf{r}}')+{\mathbf{b}}({\mathbf{r}}')))
\cdot({\mathbf{s}}-{\mathbf{r}}')}
{|{\mathbf{s}}-{\mathbf{r}}'|^3}dV'\nonumber\\
&&-\frac{\lambda}{4\pi}\int_V\frac{{\mathbf{A}}({\mathbf{r}}')
\cdot({\mathbf{s}}-{\mathbf{r}}')}{|{\mathbf{s}}-{\mathbf{r}}'|^3}dV'-
\frac{1}{4\pi}\int_S\phi({\mathbf{s}}'){\mathbf{n}}({\mathbf{s}}')
\cdot\frac{{\mathbf{s}}-{\mathbf{s}}'}{|{\mathbf{s}}-{\mathbf{s}}'|^3}dS'  \label{eq5}\\
{\mathbf{A}}({\mathbf{r}})&=&\frac{1}{4\pi}\int_V
\frac{({\mathbf{B}}_0({\mathbf{r}}')+{\mathbf{b}}({\mathbf{r}}'))
\times({\mathbf{r}}-{\mathbf{r}}')}{|{\mathbf{r}}-{\mathbf{r}}'|^3}dV' \nonumber \\
&&+\frac{1}{4\pi}\int_S{\mathbf{n}}({\mathbf{s}}')\times
\frac{{\mathbf{B}}_0({\mathbf{s}}')+{\mathbf{b}}({\mathbf{s}}')}{|{\mathbf{r}}-{\mathbf{s}}'|}dS',  \label{eq6}
\end{eqnarray}
where $\mu$ is the permeability of the fluid.

Note that this integral equation system is by far not the only possible one.
The double use of the magnetic field and its vector potential
might even seem a bit awkward. There are indeed other possible schemes, 
one of them starting from the Helmholtz equation for the vector potential 
which leads, however, to
a nonlinear eigenvalue problem in $\lambda$ (see Dobler and R\"adler 1998),
while the above scheme  is a linear eigenvalue in $\lambda$ which has many 
advantages when it comes to the numerical treatment.

The general integral equation system (\ref{eq4}-\ref{eq6}) can now 
be further specified.
In the case with ${\bf B}_0 \ne 0$ and small $R_m$, i.e. below the
threshold of self-excitation, the system describes an
induction problem in which the applied field ${\bf B}_0$ is only slightly deformed
by the velocity. In the case of ${\bf B}_0 = 0$, the system represents an eigenvalue
problem for the complex constant $\lambda$, with only negative eigenvalues
for small $R_m$ (below the dynamo threshold), and one or 
a few eigenvalues with positive real parts
for the case of large $R_m$ (above the dynamo threshold). 
 
In the case of a steady problem, with neither any growth/decay nor any oscillation
of the field,
this equation system reduces to 

\begin{eqnarray}
{\mathbf{b}}({\mathbf{r}})&=&\frac{\mu\sigma}{4\pi}\int_V\frac{({\mathbf{u}}({\mathbf{r}}')
\times({\mathbf{B}}_0({\mathbf{r}}')+{\mathbf{b}}({\mathbf{r}}')))
\times({\mathbf{r}}-{\mathbf{r}}')}
{|{\mathbf{r}}-{\mathbf{r}}'|^3}dV' \nonumber\\
&&-\frac{\mu\sigma}{4\pi}\int_S\phi({\mathbf{s}}'){\mathbf{n}}({\mathbf{s}}')\times
\frac{{\mathbf{r}}-{\mathbf{s}}'}{|{\mathbf{r}}-{\mathbf{s}}'|^3}dS' \label{eq7}\\
\frac{1}{2}\phi({\mathbf{s}})&=&\frac{1}{4\pi}\int_V\frac{({\mathbf{u}}({\mathbf{r}}')
\times({\mathbf{B}}_0({\mathbf{r}}')+{\mathbf{b}}({\mathbf{r}}')))\cdot({\mathbf{s}}-{\mathbf{r}}')}
{|{\mathbf{s}}-{\mathbf{r}}'|^3}dV'\nonumber\\
&&-\frac{1}{4\pi}\int_S\phi({\mathbf{s}}'){\mathbf{n}}({\mathbf{s}}')
\cdot\frac{{\mathbf{s}}-{\mathbf{s}}'}{|{\mathbf{s}}-{\mathbf{s}}'|^3}dS' . \label{eq8}
\end{eqnarray}

Again, we can distinguish the case with ${\bf B}_0 \ne 0$ and small $R_m$ for 
which it represent a stationary induction problem. This is of particular
interest for many technical liquid metal problems (steel casting, crystal growth) 
characterized by a rather small $R_m$. In this case,
the induced field $\bf b$ can be omitted under the integrals which leads to
a linear relation between the source $\bf u$ and the result $\bf b$
of the induction process. This linear relationship represents a convenient starting point
for the treatment  of the inverse problem to infer $\bf u$ in the bulk of the fluid from
values of $\bf b$ measured in the exteriour of the fluid.
A problem of this sort will be discussed in the next but one section.
In the case of ${\bf B}_0 = 0$, the steady system represents again an 
eigenvalue problem, but now for the critical $R_m$ that leads 
to a marginal and non-oscillatory dynamo.

\section{Numerical implementation and dimensional reductions}

In this section we will illustrate the general method by treating 
dynamo problems first in 3D and then in 2D and 1D, 
the latter cases being characterized by different degrees of 
symmetries that can be exploited for 
a dimensional reduction of the problem. We will see that, in 
stark contrast to the
corresponding procedure in the differential equation approach, such a dimensional
reduction represents a formidable task in the IEA.

\subsection{3D - Matchbox dynamos}
Let us start with a full problem in 3D for which we will
summarize the result of the papers (Xu {\it et al.} 2004a, 2004b) 
for a so-called ''matchbox dynamo'', i.e. a mean-field dynamo in a 
rectangular box.
The original assumption of a 
full velocity field as the source of induction can be easily 
translated to the case with a helical turbulence parameter $\alpha$. 
Assume that we apply a specific spatial discretization 
of all fields in Eqs. (\ref{eq4}-\ref{eq6}), we formally obtain
\begin{eqnarray}
b_i&=&\mu\sigma[P_{ik}(B_{0k}+b_k)-\lambda R_{ij}A_j-Q_{il}\phi_l], \label{eq9}\\
G_{ml}\phi_l&=&S_{mk}(B_{0k}+b_k)-\lambda T_{mj}A_j, \label{eq10}\\
A_j&=&W_{jk}(B_{0k}+b_k),\label{eq11}
\end{eqnarray}
where Einstein's summation convention is 
assumed. We have used the notion
$G_{ml}=0.5  \, \delta_{ml}+U_{ml}$. $B_{0k}$ and $b_k$
denote the degrees of freedom of the externally 
added magnetic field and the induced magnetic field, $A_j$ 
the degrees of freedom of the vector potential in the volume
$V$, $\phi_l$ the degrees of freedom of the electric potential 
at the boundary surface. For the later numerical utilization with
different induction sources it is useful that only the matrices 
$P_{ik}$ and $S_{mk}$ depend on the velocity (or on 
the corresponding mean-field induction
sources),
while $R_{ij}$, $Q_{il}$, $T_{mj}$, $G_{ml}$ and $W_{jk}$ 
depend only on the geometry of the dynamo domain and on the 
details of the discretization.

Inserting Eqs. (\ref{eq10}) and (\ref{eq11}) into Eq. (\ref{eq9}), which 
eliminates $A_j$ and $\phi_l$, we end up with one single matrix equation
for the induced magnetic field components $b_i$:
\begin{eqnarray}{\label{eq12}}
b_i&=&\mu\sigma[P_{ik}(B_{0k}+b_k)-\lambda 
R_{ij} W_{jk}(B_{0k}+b_k)-Q_{il}G_{lm}^{-1}S_{mk}(B_{0k}+b_k)\nonumber\\
&&+\lambda Q_{il}G_{lm}^{-1}T_{mj}W_{jk}(B_{0k}+b_k)]
\end{eqnarray}
which can be further transformed into
\begin{eqnarray}{\label{eq13}}
[\delta_{ik}-\mu\sigma E_{ik}-\mu\sigma\lambda F_{ik}]b_k=[\mu\sigma E_{ik}+
\mu\sigma\lambda F_{ik}]B_{0k},
\end{eqnarray}
where $E_{ik}=P_{ik}-Q_{il}G_{lm}^{-1}S_{mk}$ and                                      %!!!!
$F_{ik}=-R_{ij}W_{jk}+Q_{il}G_{lm}^{-1}T_{mj}W_{jk}$.
It should be noted, though, that the needed inversion of the boundary integral
equation (\ref{eq10}) for the electric potential needs some specific care
(Wondrak {\it et al.} 2009).
This has to do with the singular character of this equation which, in turn,
mirrors the ambiguity of the electric potential with respect to some additive
constant. 

For the computation of induction effects, the induced magnetic field can be 
obtained by solving the algebraic equation system (\ref{eq12}). 
In order to compute kinematic dynamos without external sources, Eq.(\ref{eq12}) 
reduces to the following generalized eigenvalue problem for $\lambda$: 
\begin{eqnarray}{\label{eq14}}
[\delta_{ik}-\mu \sigma E_{ik}]b_k=\lambda \mu \sigma F_{ik}b_k,                               %!!!!
\end{eqnarray}

To test the working of the scheme we apply it now to  $\alpha^2$ 
dynamos in rectangular geometry, which we would like to coin ''matchbox dynamos''. 
As usual in mean-field dynamo theory (see Krause 
and R\"adler 1980), $\alpha$ parametrizes the induction effect of helical turbulence
on the large scale magnetic field. Formally speaking, the term ${\bf u} \times {\bf B}$ in the
electromotive force term in Eq. (\ref{eq3}) has to be replaced by $\alpha   \cdot {\bf B}$.

While the spherically symmetric $\alpha^2$ dynamo has played a 
paradigmatic role in 
dynamo theory, its  counterpart in rectangular geometry is certainly a 
highly artificial problem
without any application in astrophysics. Nevertheless, it may 
illustrate the capabilities of the IEA to solve dynamo problems
in geometries that are far away from spherical.

In our specific example (for more cases see Xu {\it et al.} 2004b) we 
consider a homogeneous distribution, $\alpha({\bf r})=C$, in a 
''matchbox'' with side lengths 2.0x1.6x1.2 which gives 
approximately the same volume  as a corresponding sphere of radius 1.
The domain is divided into $10^3$ smaller rectangular boxes, with each face
divided into $10^2$ rectangles.

\begin{figure}[h!]
\begin{center}
\epsfxsize=14cm{\epsffile{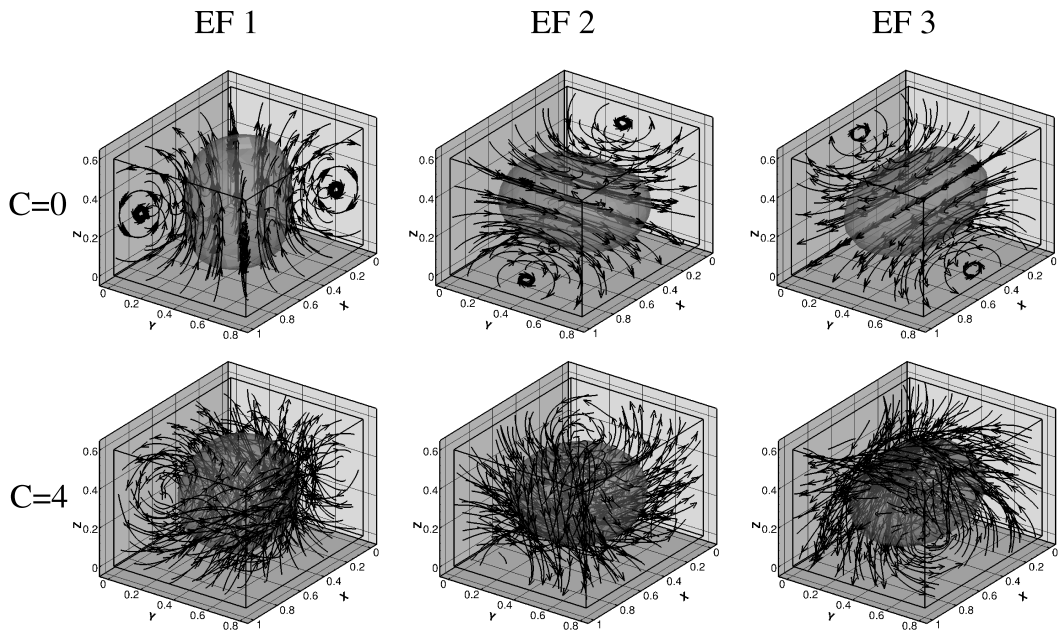}}
\end{center}
\caption{Magnetic eigenfields corresponding to the three dominant
eigenvalues of the $\alpha^2$ dynamo in a matchbox with sidelengthes ratio 2x1.6x1.2. }
\end{figure}

For this setting, we find the first three critical values of $C$ to be
$C'_1=4.728$,  $C'_2=4.898$, and $C'_3=4.934$. The corresponding first three
eigenfunctions are shown, for the free decay case $C=0$ and the slightly 
sub-critical case $C=4$, in figure 1. These eigenfunctions still resemble the
corresponding eigenfunctions in the spherical case. However, the exact degeneracy of the
eigenvalues in the latter case is lifted now. The lowest eigenvalue
corresponds to an eigenfunction whose dipolar axis is perpendicular to the
largest face. Nevertheless, the three eigenvalues are still close the well-known 
value 4.49 for the spherical case.

\subsection{2D - VKS-like dynamos}

The recent successful dynamo experiments (see Stefani {\it et al.} 2008 for
a recent survey)
in Riga (Gailitis {\it et al.} 2000), 
Karlsruhe
(Stieglitz and M\"uller 2001), and Cadarache (Monchaux {\it et al.} 2009) 
were all carried out in cylindrical vessels filled with liquid sodium. 
For this reason it is worth to specify the integral equation approach 
to this geometry.
As long as the dynamo source (i.e. the velocity field
or a corresponding mean-field
quantity) is axisymmetric, the different azimuthal modes of the electromagnetic fields 
with the 
angular dependence according to $\exp(i m \varphi)$
can be decoupled. Ultimately, this leads to a tremendous
reduction of the numerical effort. The price we have to pay for this
is the necessity to carefully deriving the dimensionally reduced
version of the integral equation system. The necessary integrations over
$\varphi$  
turnes out to be a formidable task (see Xu {\it et al.} 2008), 
quite in contrast to the respective 
procedure in the differential equation approach, where the expression 
$\partial/\partial \varphi$ can simply replaced by $i m$.

Lets assume now the electrically
conducting fluid to be confined in a cylinder
with radius $R$ and total height $2H$. 
Introducing the cylindrical coordinate system ($\rho, \varphi, z$), we have
\begin{eqnarray}
{\mathbf{r}}=[\rho \cos \varphi, \rho \sin \varphi, z]^T,
{\mathbf{b}}=[b_\rho, b_\varphi, b_z]^T,
{\mathbf{u}}=[u_\rho, u_\varphi, u_z]^T.\label{eq14a}
\end{eqnarray}
The magnetic field ${\mathbf{b}}$, the electric potential $\phi$, and 
the vector potential ${\mathbf{A}}$ all are expanded into azimuthal modes:
\begin{eqnarray}{\label{eq15}}
\pmatrix{{\mathbf{b}}\cr
\phi\cr
{\mathbf{A}}}=\sum_{m=-\infty}^{\infty} \pmatrix{{\mathbf{b}}_m\cr 
\phi_m\cr {\mathbf{A}}_m} \exp(im\varphi).
\end{eqnarray}
As long as  the velocity field is axisymmetric
(i.e. it has only a component with
$m=0$), the fields $[{\mathbf{b}}_m, \phi_m,{\mathbf{A}}_m]^T$                            %!!!!
with different $m=0, \pm 1, \pm 2,\cdots$ decouple from each other. 
For the sake of convenience,                                                                               %!!!
we will always re-denote $[{\mathbf{b}}_m, \phi_m, {\mathbf{A}}_m]^T$                                     %!!!
as $[{\mathbf{b}}, \phi, {\mathbf{A}}]^T$.                    %!!! 
The reduction to a problem exclusively in $r$ and $z$ is then achieved
by carrying out all integrations over $\phi$. This painstaking procedure results 
finally in  the following system of matrix equations

\begin{eqnarray}
\pmatrix{{{b}}_\rho\cr 
{{b}}_\varphi\cr
{{b}}_z\cr}&=&\mu\sigma \left[  {\mathbf{P}}\pmatrix{  {{B}}_{0\rho}+{{b}}_\rho   \cr
{{B}}_{0\varphi}+{{b}}_\varphi              \cr
{{B}}_{0z}+b_z  }-{\mathbf{Q}}\pmatrix{{{\phi}}_{s1}\cr
{{\phi}}_{s2}\cr
{{\phi}}_{s3}}-\lambda{\mathbf{R}}\pmatrix{{{A}}_\rho\cr 
{{A}}_\varphi\cr
{{A}}_z\cr} \right],\label{eq16}\\
\frac{1}{2}\pmatrix{{\mathbf{\phi}}_{s1}\cr
{\mathbf{\phi}}_{s2}\cr
{\mathbf{\phi}}_{s3}}&=&{\mathbf{S}}\pmatrix{
{{B}}_{0\rho}+{{b}}_\rho   \cr 
{{B}}_{0\varphi}+{{b}}_\varphi              \cr
{{B}}_{0z}+b_z
}-\lambda{\mathbf{T}}\pmatrix{{{A}}_\rho\cr 
{{A}}_\varphi\cr
{{A}}_z}-{\mathbf{U}}\pmatrix{{\mathbf{\phi}}_{s1}\cr
{{\phi}}_{s2}\cr
{{\phi}}_{s3}},\label{eq17}\\
\pmatrix{{{A}}_\rho\cr 
{{A}}_\varphi\cr
{{A}}_z}&=&{\mathbf{W}}\pmatrix{
{{B}}_{0\rho}+{{b}}_\rho   \cr 
{{B}}_{0\varphi}+{{b}}_\varphi              \cr
{{B}}_{0z}+b_z
   \cr},\label{eq18}
\end{eqnarray}
where the matrix elements of $\mathbf{P}$, $\mathbf{Q}$, $\mathbf{R}$, $\mathbf{S}$, $\mathbf{T}$,
$\mathbf{U}$, and $\mathbf{W}$ can be found in the paper by  Xu {\it et al.} (2008).
Combining Eqs.(\ref{eq16}-\ref{eq18}), we obtain
\begin{eqnarray}{\label{eq19}}
({\mathbf{I}}-\mu\sigma {\mathbf{E}}-\mu\sigma\lambda{\mathbf{F}}){\mathbf{b}}=\mu\sigma({\mathbf{E}}+\lambda{\mathbf{F}}){\mathbf{B}}_0,
\end{eqnarray}
where
\begin{eqnarray}
{\mathbf{E}}&=&{\mathbf{P}}-{\mathbf{Q}} \cdot(\frac{1}{2}{\mathbf{I}}+{\mathbf{U}})^{-1} 
\cdot {\mathbf{S}}  \label{eq20}\\
{\mathbf{F}}&=&{\mathbf{Q}} \cdot (\frac{1}{2}{\mathbf{I}} + 
{\mathbf{U}})^{-1} \cdot {\mathbf{T}} \cdot {\mathbf{W}} - {\mathbf{R}} \cdot {\mathbf{W}} \; .\label{eq21}
\end{eqnarray}
Again, induced magnetic fields can be obtained by 
solving the algebraic equation system (\ref{eq19}), while
for kinematic dynamo problems, the generalized eigenvalue problem
\begin{eqnarray}{\label{eq22}}
({\mathbf{I}}-\mu \sigma {\mathbf{E}})\cdot{\mathbf{b}}=\mu \sigma \lambda {\mathbf{F}} \cdot {\mathbf{b}}
\end{eqnarray}
has to be solved. 

In connection with the optimization of the VKS dynamo experiment,
Mari\'{e}, Normand and Daviaud (2006)  had studied an analytical 
test flow of the same topological type as the flow in the real experiment 
(flow topology s2$^+$t2, what means two poloidal eddies 
with radial inflow
in the equatorial plane, together with two counter-rotating toroidal eddies).
The velocity field of this ''MND flow'' reads:
\begin{eqnarray}
u_r&=&-\frac{\pi}{2} \; r \; (1-r)^2 (1+2r) \cos (\pi z)\label{eq23}\\
u_{\varphi}&=&4 \epsilon r (1-r) \sin (\pi z/2)\label{eq24}\\
u_z&=&(1-r)(1+r-5 r^2) \sin (\pi z) . \label{eq25}
\end{eqnarray}
For the parameter $\epsilon$, which determines the ratio of toroidal to poloidal 
flow, we have used the value $\epsilon=0.7259$. For 
the case 
without any side layer, the structure of the magnetic eigenfield is illustrated 
in figure 2. In figure 2a the field lines of the equatorial dipole are seen,
while the isosurfaces  of 
the magnetic field energy (figure 2b) show the typical banana-like structure. 

\begin{figure}[h!]
\begin{center}
\epsfxsize=14cm{\epsffile{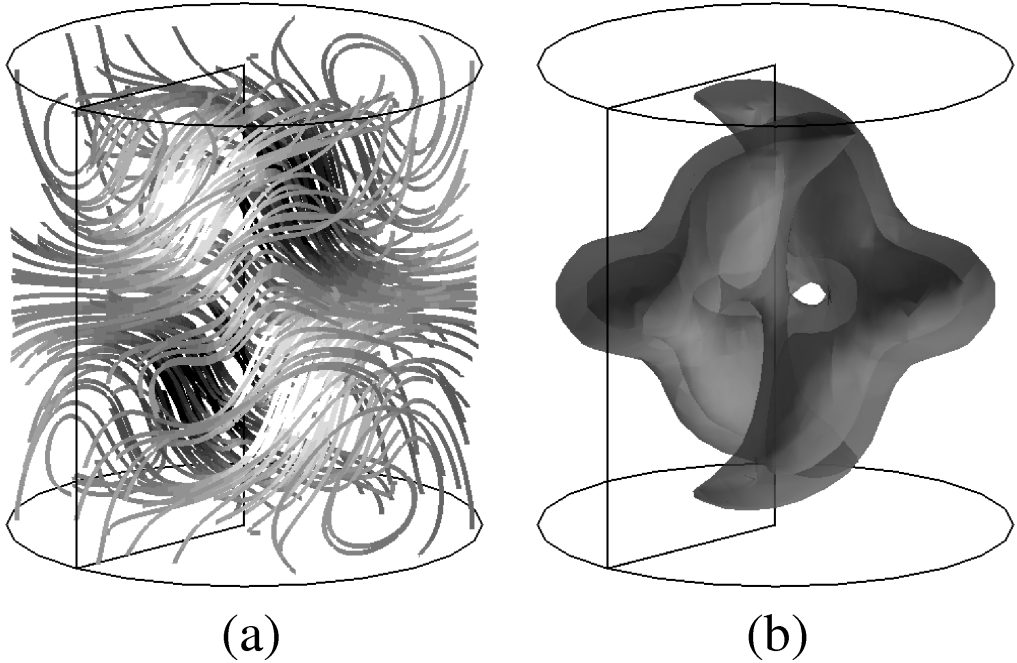}}
\end{center}
\caption{Simulated magnetic field structure of the dominant eigenmode of
the MND flow. 
(a) Magnetic field lines of the equatorial dipole field. (b) Iso-surface of the 
magnetic field energy.}
\end{figure}

\subsection{Reduction to 1D - Spherically symmetric $\alpha^2$ dynamo}

After having utilized, in the former subsection, the cylindrical symmetry of some 
flow field for  reducing the full 3D IEA system to a 2D system, we go now
one step further and invoke spherical symmetry to make a reduction to a 1D 
problem. This is possible for a spherically symmetric $\alpha^2$ dynamo.
In contrast to the original, analytically solvable model with constant 
$\alpha$ (Krause and R\"adler 1980),
we will allow the profile $\alpha$ to  vary with the radial coordinate $r$.
After summarizing the derivation of the two coupled radial integral equations
as given in (Stefani {\it et al.} 2000, Xu {\it et al.} 2004b), we will
treat numerically a non-trivial model with a radial dependence $\alpha(r)=C r^2$.

The divergence-free magnetic field
$\bf{B}$  is split, as usual in dynamo theory, 
into a poloidal and a toroidal part, denoted by ${\bf{B}}_{P}$ and
${\bf{B}}_{T}$. Using the Coulomb gauge, $\nabla \cdot {\bf A}=0$, the same can be 
done for the vector potential, 
${\bf A}={\bf{A}}_{P}+{\bf{A}}_{T}$. 
All these fields can be represented by the defining scalars $S$, $T$, $S^A$, $T^A$
according to  
\begin{eqnarray}{\label{eq26}}
{\bf{B}}_{P}=\nabla \times \nabla \times\left(\frac{S}{r} \, {\bf{r}} 
\right),
\; \; \; 
{\bf{B}}_{T}=\nabla\times \left(\frac{T}{r} \, {\bf{r}} \right) \; ,
\end{eqnarray}
\begin{eqnarray}{\label{eq27}}
{\bf{A}}_{P}=\nabla \times \nabla \times\left(\frac{S^A}{r} \, {\bf{r}} 
\right),
\; \; \; 
{\bf{A}}_{T}=\nabla\times \left(\frac{T^A}{r} \, {\bf{r}} \right) \; .
\end{eqnarray}
For our spherical problem, 
the defining scalars and the electric potential 
can be expanded in series of spherical harmonics $Y_{lm}(\theta,\phi)$. 
As an example we indicate 
\begin{eqnarray}{\label{eq28}}
S(r,\theta,\phi)&=&\sum_{l,m} s_{lm}(r) Y_{lm}(\theta,\phi) \; ,
\end{eqnarray}
and corresponding expressions can be written for $T(r,\theta,\phi)$, $S^A(r,\theta,\phi)$, 
$T^A(r,\theta,\phi)$ and $\varphi(r,\theta,\phi)$, with   
$s_{lm}(r)$ being replaced by $t_{lm}(r)$, $s^A_{lm}(r)$, $t^A_{lm}(r)$, and 
$\varphi_{lm}(r)$, respectively.

For the spherical harmonics $Y_{lm}(\theta,\phi)$ 
the definition
\begin{eqnarray}{\label{eq29}}
Y_{lm}(\theta,\phi)=\sqrt{\frac{2l+1}{4 \pi}\frac{(l-m)!}{(l+m)!}} \; 
P_{lm}(\cos \theta) e^{im \phi} \
\end{eqnarray}
is employed, which implies
the following orthogonality relation for the $Y_{lm}(\theta,\phi)$:
\begin{eqnarray}{\label{eq30}}
\int_0^{2\pi} d\phi \int_0^{\pi} \sin \theta \, d 
\theta \, Y_{l'm'}^{\ast}(\theta,\phi)
Y_{lm}(\theta,\phi)=\delta_{ll'} \delta_{mm'} \; .
\end{eqnarray}

From Eqs. (\ref{eq26}-\ref{eq28})  we obtain the components of 
$\bf{B}$ in the form
\begin{eqnarray}
B_r(r,\theta,\phi)&=&\sum\limits_{l,m} \frac{l(l+1)}{r^2} s_{lm} (r) 
Y_{lm}(\theta,\phi)   \label{eq31} \\
B_{\theta}(r,\theta,\phi)&=&\sum\limits_{l,m} \left(
\frac{t_{lm}(r)}{r \sin \theta} \frac{\partial Y_{lm}(\theta,\phi)}
{\partial \phi}+
\frac{1}{r} \frac{d s_{lm} (r)}{d r}
\frac{\partial Y_{lm}(\theta,\phi)}
{\partial \theta} \right)  \label{eq32}\\
B_{\phi}(r,\theta,\phi)&=&\sum\limits_{l,m} \left(  
- \frac{t_{lm}(r)}{r} \frac{\partial Y_{lm}(\theta,\phi)}
{\partial \theta}+\frac{1}{r \sin \theta} \frac{d s_{lm} (r)}{d r}
\frac{\partial Y_{lm}(\theta,\phi)}
{\partial \phi} \right) \; , \label{eq33}
\end{eqnarray}
and  equivalent expressions for the components of $\bf A$, in which 
$s_{lm}(r)$ and $t_{lm}(r)$ are replaced by $s^A_{lm} (r)$ and $t^A_{lm} (r)$, 
respectively. 

A necessary ingredient for the dimensional reduction is 
the expression for the inverse distance
between two points $\bf{r}$ and $\bf{r'}$ in terms of spherical harmonics, 
\begin{eqnarray}{\label{eq34}}
\frac{1}{|{\bf{r}}-{\bf{r'}}|}=4 \pi \sum_{l=0}^{\infty} 
\sum_{m=-l}^{l}
\frac{1}{2l+1} \frac{r_{<}^l}{r_{>}^{l+1}} 
Y_{lm}^{\ast}(\theta',\phi')
Y_{lm}(\theta,\phi) \; ,
\end{eqnarray}
where $r_{>}$ denotes the larger of the values $r$ and $r'$, and
$r_{<}$  the smaller one. 

On this basis, the two 
coupled integral equations 
for the functions $s_{lm}(r)$ and 
$t_{lm}(r)$ can be derived. 
The first equation for $s_{lm}(r)$ can easily be obtained by multiplying the
magnetic field with the unit vector in radial direction and then integrating over
the angles $\theta$ and $\phi$. The derivation of the equation for
$t_{lm}(r)$ needs more work, including the treatment of the electric potential
and the vector potential at the boundary.

Here we go straight to the final result of this procedure, whose details were worked out  
by Xu {\it et al} (2004b): 
$s_{lm}(r)$ and $t_{lm}(r)$:
\begin{eqnarray}{\label{eq35}}
s_{lm}(r)&=&\frac{\mu_0 \sigma }{2l+1} \left[
\int_0^r  \frac{{r'}^{l+1}}{r^{l}} \, \alpha(r') \, t_{lm}(r') 
\, dr'+
\int_r^R  \frac{r^{l+1}}{{r'}^{l}} \, \alpha(r') \, t_{lm}(r') \, 
dr' \right. \nonumber\\
&&\left. -\lambda\int_0^r\frac{{r'}^{l+1}}{r^l}s_{lm}(r')dr'-\lambda\int_r^R
\frac{r^{l+1}}{{r'}^l}s_{lm}(r')dr' \right] \; .\
\end{eqnarray}
and
\begin{eqnarray}{\label{eq36}}
t_{lm}(r)&=&\mu_0\sigma \left[ \alpha(r)s_{lm}(r)-\frac{l+1}{2l+1}\int_0^r\frac{d\alpha(r')}
{dr'}s_{lm}(r')\frac{{r'}^l}{r^l}dr'\label{eq41} \right. \nonumber\\
&&+\frac{l}{2l+1}\int_r^R\frac{d\alpha(r')}{dr'}s_{lm}(r')\frac{r^{l+1}}
{{r'}^{l+1}}dr'+
\frac{l+1}{2l+1}\frac{r^{l+1}}{R^{2l+1}}\int_0^R{r'}^l
\frac{d\alpha(r')}{dr'}s_{lm}(r')dr'\nonumber\\
&&+\frac{\lambda }{2l+1}\frac{r^{l+1}}{R^{2l+1}}\int_0^R{r'}^{l+1}t_{lm}(r')dr'-
\frac{\lambda}{2l+1}\int_0^r\frac{{r'}^{l+1}}{r^l}t_{lm}(r')dr'\nonumber\\
&&\left. -\frac{\lambda}{2l+1}\int_r^R\frac{r^{l+1}}{{r'}^l}t_{lm}(r')dr'-
\frac{r^{l+1}}{R^{l+1}}
\alpha(R)s_{lm}(R)] \right] \; .
\end{eqnarray}

The correctness of this system can be easily checked by differentiating
equations (\ref{eq35}) and (\ref{eq36}) two times with respect to the radius 
which gives the two well-known differential equations
for $s_{lm}(r)$ and $t_{lm}(r)$:
\begin{eqnarray}{\label{eq37}}
\lambda s_{lm}=\frac{1}{\mu_0\sigma} \left[ \frac{d^2s_{lm}}{dr^2}-
\frac{l(l+1)}{r^2}s_{lm} \right] +
\alpha(r)t_{lm},
\end{eqnarray}
\begin{eqnarray}{\label{eq38}}
\lambda t_{lm}=\frac{1}{\mu_0\sigma}\left[ \frac{d^2t_{lm}}{dr^2}-
\frac{l(l+1)}{r^2}t_{lm} \right]-
\frac{d}{dr}\left( \alpha(r)\frac{ds_{lm}}{dr} \right)+\frac{l(l+1)}{r^2}\alpha(r)s_{lm},
\end{eqnarray}
with the boundary conditions
\begin{eqnarray}{\label{eq39}}
\left. t_{lm}(R)=R\frac{ds_{lm}(r)}{dr} \right|_{r=R}+l s_{lm}(R)=0.
\end{eqnarray}

It should be noted that in this particular case the complicated dimensional
reduction (by expressing the fields in spherical harmonics and integrating over
the spherical angles) can be circumvented by a simpler one. This starts 
directly from the
differential equation system (\ref{eq37}-\ref{eq39}) for which the 
Green functions can be derived which then lead to the 
integral equation system (\ref{eq35}-\ref{eq36}) (Xu {\it et al.}2004b).

Now we turn to the numerical utilization of the radial integral 
equation system (\ref{eq35}-\ref{eq36}). The discretization in radial
direction was done with a grid number of 128 which gives already a 
satisfactory accuracy. The eigenvalue solver
gives us immediately all the eigenvalues. This is illustrated in figure 3, which
shows for the case $\alpha(r)=C r^2$ the real part (figure 3a) and the imaginary
part (figure 3b) of the first 8 eigenvalues. The existence of two so-called
exceptional points, at which two real eigenvalue branches coalesce and 
continue as a pair of complex conjugated eigenvalues, is clearly seen.

\begin{figure}[h!]
\begin{center}
\epsfxsize=10cm{\epsffile{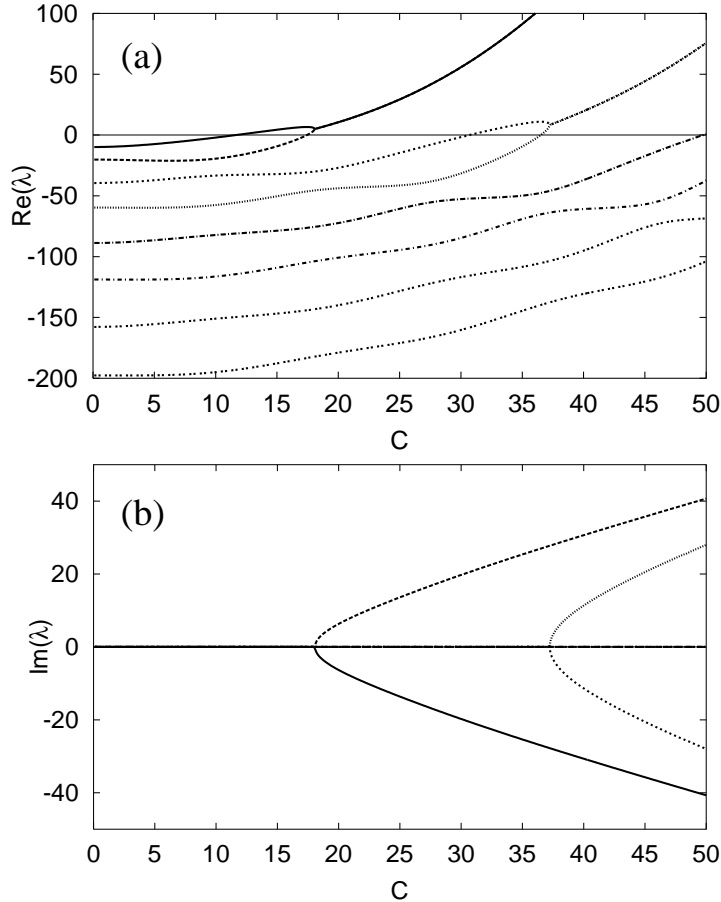}}
\end{center}
\caption{Eigenvalues of the spherically symmetric $\alpha$ dynamo with
a radial dependence $\alpha(r)=C r^2$. (a) Growth rates. (b) Frequencies.}
\end{figure}

\section{Inverse problems for small $R_m$}

Assuming a velocity field as given, and asking for the 
resulting magnetic field, represents a typical forward problem. 
The question can, however,
also be inverted: assume the magnetic field as being 
measured in the exteriour of the fluid, 
what is the velocity that produces this field? This is a typical
inverse problem, and it has puzzled dynamo theorist since many years.

In the general case, for arbitrary values of $R_m$, it represents a 
highly nonlinear inverse problem.  Some progress in its treatment has been made
for restricted set-ups, for example by applying the so-called frozen-flux
approximation for the Earth's core which allows to determine (still
with some appropriate regularization) solutions for the  
tangential velocity field components at the Core-Mantel boundary 
from the measured time dependence of the
radial magnetic field components (Roberts and Scott 1965). 
A complementary sort of restricted models
is concerned with  the determination of the radial dependence of $\alpha(r)$
of an assumed spherically symmetric $\alpha^2$ model from spectral 
properties. Solving such a type of restricted inverse problems (by means of 
an Evolutionary Strategy) it was possible, e.g., 
to obtain such $\alpha$ profiles that lead to oscillatory dynamo solutions
(Stefani and Gerbeth 2003).
Apart from those special solutions, inverse dynamo theory is still
in its infancy.

The situation becomes much better for the case of induction problems 
with  small $R_m$. As mentioned above
the induced field $\bf b$ can then be omitted under the integrals,
and we obtain a linear relation between the (wanted) source $\bf u$ and the 
(measured) result 
$\bf b$ of the induction process. 

Although it is far from being a generic geo- or astrophysical problem, we would like
to illustrate the typical solution procedure of such inverse problems by a model related
to industrial steel casting. Figure 4a shows the set-up 
of  a corresponding model experiment in which a liquid metal 
(in our case GaInSn) is poured from a tundish through an submerged entry nozzle 
into the mould. Among other problems of industrial interest, like flow instabilities
due to Argon entrainment into the metal, we deal here with the problem how
a magnetic stirrer can influence the flow structure in the mould. For this purpose, we use
a so-called one-port submerged entry nozzle with just a hole at the bottom of the nozzle.

Based on the integral equation system (\ref{eq7}-\ref{eq8}), specified to small $R_m$, we
have developed  the so-called Contactless Inductive Flow Tomography (CIFT)
for the reconstruction of velocity fields from externally measured magnetic fields 
(Stefani and Gerbeth 1999, 2000a, b, Stefani et al. 2004).
While CIFT is in principle able to infer full 3D velocity fields by applying subsequently
two different (e.g. orthogonal) external magnetic fields, for the present case of
thin slab casting it can be reduced to the determination of the
velocity components parallel to the wide faces of the mould (Wondrak {\it et al.} 2010).
For this it is enough to apply only one magnetic field by  a single coil (see figure 4a).
The interaction of the flow with the applied field produces 
induced magnetic fields that we measure at a number of positions at the narrow
faces of the mould in order to reconstruct from them the velocity field.
The mathematics of this inversion relies in the minimization of the mean squared 
deviation of the measured magnetic fields from the fields resulting according to
the integral equation system (\ref{eq7}-\ref{eq8}) from the velocity field. 
This minimization is done by solving the normal 
equations, whereby we use various auxiliary functionals which serve to 
ensure the divergence-free condition of the velocity,
to enforce its two-dimensionality, and to minimize in parallel its mean quadratic
value, weighted by some regularization parameter (Tikhonov regularization).
Figure 4b shows a time sequence of reconstructed velocity fields resulting from this 
inversion, for a particular flow experiment with applied stirring. 
What is clearly seen from the four  plots is a significant (left-right asymmetric) 
up- and down movement of the vortex centers (indicated by the dark patches).

\begin{figure}[h!]
\begin{center}
\epsfxsize=14cm{\epsffile{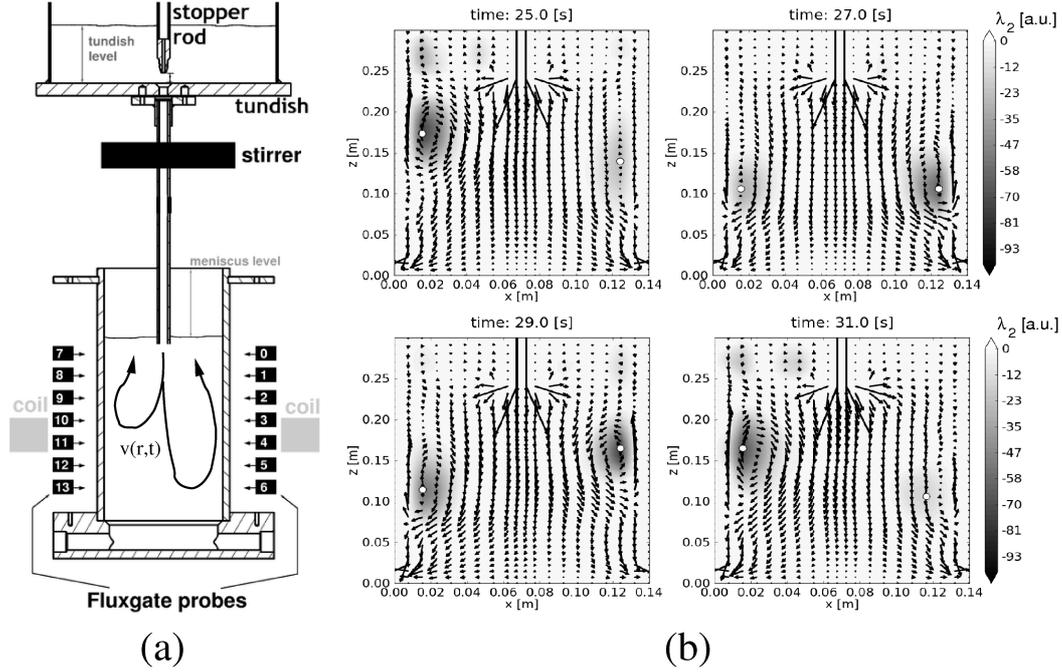}}
\caption{Illustration of the CIFT technique at a model of continuous casting.
(a) Experimental set-up with the tundish, the single-port submerged 
entry nozzle, and the mould.
The external field is produced by a coil, and the 
induced fields are measured
at 7 fluxgate sensors at each narrow face of the mould. 
(b) Four subsequent plots of the velocity field as reconstructed from
the measured induced magnetic fields. The quantity $\lambda_2$, indicated 
by the grey scale, 
is the most appropriate measure for 
the identification of a vortex (Jeong and Hussain 1995).}
\end{center}
\end{figure}

\section{Conclusions}
In this paper we have surveyed the principles and 
some applications of the integral equation approach (IEA) 
to kinematic MHD. The IEA has turned out
a viable scheme for the correct numerical treatment of 
dynamo problems in non-spherical
domains, for examples in ''matchboxes'' and, more importantly, in 
cylinders (see also Avalos-Zu\~{n}iga et al. 2007). It has also 
proved a good starting point for the treatment of inverse problems of 
MHD, with direct applications
in a number of technical problems characterized by small $R_m$.

\section*{Acknowledgments}
This work was supported by Deutsche Forschungsgemeinschaft
in frame of SFB 609 and under Grant No. STE 991/1-1.
We thank Andr\'e Giesecke for many fruitful discussions on
various numerical methods in MHD.

\end{document}